\documentclass[aps,pre,twocolumn,superscriptaddress]{revtex4}

\usepackage{graphicx,epsfig}  
\usepackage{dcolumn}   
\usepackage{bm}        
\usepackage{amssymb}   
\usepackage{color}
\usepackage{mathtools}
\usepackage{esvect}
\usepackage{float}
\usepackage{makecell}

\begin{document}

\title{Coupling and Acceleration of Externally Injected Electron Beams in Laser-Driven Plasma Wakefields}

\author{Srimanta Maity}
\email {srimantamaity96@gmail.com}
\affiliation{ELI Beamlines Facility, The Extreme Light Infrastructure ERIC, Za Radnicí 835, 25241 Dolní Břežany, Czech Republic}

\author{Pavel Sasorov}
\affiliation{ELI Beamlines Facility, The Extreme Light Infrastructure ERIC, Za Radnicí 835, 25241 Dolní Břežany, Czech Republic}

\author{Alexander Molodozhentsev}
\email {alexander.molodozhentsev@eli-beams.eu}
\affiliation{ELI Beamlines Facility, The Extreme Light Infrastructure ERIC, Za Radnicí 835, 25241 Dolní Břežany, Czech Republic}

\begin{abstract}

\noindent 
 The multi-stage method of laser wakefield acceleration (LWFA) presents a promising approach for developing stable, full-optical, high-energy electron accelerators. By segmenting the acceleration process into several booster stages, each powered by independent laser drivers, this technique effectively mitigates challenges such as electron dephasing, pump depletion, and laser diffraction. A critical aspect of multi-stage LWFA is the nonlinear interaction between the injected electron beam and the laser-driven wakefields in the booster stage. This study investigates the injection and acceleration of external electron beams within wakefields in the booster stage using multi-dimensional Particle-In-Cell (PIC) simulations. We provide both qualitative and quantitative descriptions of the observed physical processes. Key parameters influencing charge coupling process and the resultant beam quality have been identified. Furthermore, we have examined how off-axis injection relative to the driver laser influences the acceleration process and beam quality. Our findings provide valuable insights for advancing and optimizing multi-stage plasma-based accelerators.

\end{abstract}

\maketitle        
\section{Introduction}\label{intro}

Laser Wakefield Acceleration (LWFA) \cite{PhysRevLett.43.267, joshi1984ultrahigh, 509991, RevModPhys.81.1229} has emerged as a highly promising technique for generating high-energy, high-quality electron beams. In traditional radio-frequency (RF) particle accelerators, the accelerating electric field is limited to around a hundred Megavolts per meter (MV/m) due to material breakdown constraints. In contrast, plasma-based accelerators can generate electric fields in the range of hundreds of Gigavolts per meter (GV/m) \cite{PhysRevLett.70.37, everett1994trapped, modena1995electron}, which is at least three orders of magnitude higher than conventional RF accelerators. This makes plasma-based accelerators, in principle, much more compact and potentially far more cost-effective. Laser-plasma-based acceleration occurs when a high-intensity laser pulse travels through a plasma medium, exerting a ponderomotive force that pushes plasma electrons away from its path, resulting in charge separation. This charge separation creates plasma density waves with intense electromagnetic fields, known as wakefields, with a phase velocity same as the group velocity of the laser pulse inside the plasma medium. These wakefields can accelerate as well as focus a trailing charged particle bunch.

Over the past two decades, a significant progress has been made in the development of laser-driven plasma accelerators \cite{malka2005laser, joshi2007development, hooker2013developments, albert20212020, kurz2021demonstration, irshad2024pareto}. Experimental studies have demonstrated the production of high-quality, quasi-monoenergetic electron beams using tabletop laser-plasma setups \cite{mangles2004monoenergetic, faure2004laser, geddes2004high, wang2021free, oubrerie2022controlled, zgadzaj2024plasma}. Quasi-monoenergetic electron beams with an energy of 8.0 GeV have been generated in a laser wakefield acceleration (LWFA) experiment utilizing a centimeter-scale capillary plasma channel \cite{PhysRevLett.122.084801}. Recent experiments \cite{aniculaesei2024acceleration, picksley2024matched} have produced electron beams with energies nearing 10 GeV, marking the highest energy achieved in a single-stage LWFA to date. To meet the demands of many applications \cite{walker2017horizon, emma2021free, vishnyakov2023compact, molodozhentsev:fls2023-th2c2, Whitehead:2024uco, Whitehead:2024dkn, galletti2024prospects} and ensure stable, high-repetition-rate operation in laser-plasma-based accelerators capable of producing high-energy, high-quality electron beams, it is crucial to improve the control over both the injection mechanism and laser-plasma parameters. 

Various injection schemes for LWFA, such as self injection \cite{bulanov1992nonlinear, xu2005electron, ohkubo2006wave, lei2023controllable, horny2024efficient}, ionization injection \cite{PhysRevLett.104.025003, PhysRevLett.104.025004, PhysRevLett.105.105003, irman2018improved, maity2024parametric}, colliding-pulse method \cite{PhysRevLett.76.2073, PhysRevLett.79.2682, faure2006controlled, wang2022injection, bohlen2023colliding}, and density down-ramp injection mechanism \cite{PhysRevE.58.R5257, PhysRevLett.100.215004, ke2021near, hue2023control} have been proposed. While these methods are effective in single-stage laser-plasma accelerators, they offer limited control over the initial properties of the injected electrons and, consequently, the quality of the resulting accelerated beam. Additionally, single-stage LWFA has inherent limitations \cite{leemans2009laser}. For example, the energy gained by injected electrons is restricted by (i) the dephasing length, which is the distance over which the electrons fall into the decelerating phase of the wakefield, and (ii) the pump depletion length, which is the distance over which the driver laser loses most of its energy and can no longer drive a strong wakefield. The diffraction of laser pulse over a long propagation distance (several Rayleigh length) is an another issue for single-stage laser-plasma-based accelerators. A multi-stage configuration of LWFA can overcome these limitations and, in principle, accelerate electron beams to unlimited energy. A multi-stage laser-plasma-based accelerator basically consists of several shorter plasma targets (booster stages) and independent laser drivers. The efficiency of a multi-stage LWFA depends on several key technical and physical factors, including the transport of the electron beam between successive booster stages, the timing synchronization between successive stages, the injection and charge coupling of the electron beam with the wakefield in each booster stage, and the maintenance of beam quality throughout the acceleration process \cite{PhysRevSTAB.3.071301}. 

Several research groups are actively working to experimentally realize multi-stage LWFA setups. For instance, two-stage laser-driven electron acceleration was demonstrated by Kimura et al. \cite{PhysRevLett.86.4041}, for the first time in a proof-of-principle experiment. In this study, the electron beam was initially generated from an conventional microwave-driven linear accelerator. They also explored the coupling between a CO$_2$ laser and an electron beam, demonstrating high trapping efficiency and narrow energy spread in a laser-driven accelerator \cite{PhysRevLett.92.054801}. The first proof-of-principle experimental study on all-optical laser wakefield acceleration in a two-stage setup was reported by Kaganovich et al. \cite{kaganovich2005first}. A dual-stage laser-wakefield acceleration scheme was also experimentally studied by Kim et al. \cite{PhysRevLett.111.165002} for the enhancement of electron energy to the multi-GeV regime. Luo et al. \cite{PhysRevLett.120.154801} demonstrated the advantages of using curved plasma channels for multi-stage LWFA, highlighting the technical benefits of this approach. Wu et al. \cite{wu2021high} experimentally investigated the external injection and subsequent acceleration of electron beams from a conventional linear accelerator into a laser-plasma accelerator. Foerster et al. \cite{foerster2022stable} reported generating stable, high-quality electron beams in a staged setup using a plasma wakefield accelerator driven by high-current electron beams from a laser wakefield accelerator. Three-stage laser wakefield accelerator driven by few-cycle laser pulses was studied by Lécz et al. \cite{lecz2023three} using Particle-In-Cell simulations. Knetsch et al. \cite{knetsch2023high} proposed a novel concept of a laser-gated multistage plasma wakefield accelerator that optimizes interplasma distances, enabling scaling to the TeV energy range with a high average accelerating gradient. Despite these advances, further research is needed to better understand the coupled nonlinear physical processes that occur in staged plasma-based accelerator setups. This is essential for the development of compact, stable multi-stage laser-plasma accelerators capable of generating high-quality electron beams in the GeV to TeV range for various applications. Thus, research on the nonlinear coupling and evolution of external electron beams in booster stages remains crucial and highly relevant.

This study explores several physical aspects of multi-stage LWFA, focusing on the interaction between externally injected electron beams and the laser-driven wakefields in a booster stage. Multi-dimensional Particle-In-Cell (PIC) simulations have been employed to investigate these processes. Previous studies \cite{jin2019coupling, pathak2020charge, pathak2021electron}, using theoretical analysis and PIC simulations, have explored charge coupling in multi-stage LWFA configurations, identifying key parameters that influence the efficient coupling and acceleration of injected electron beams in the booster stage. While our current research builds on these earlier works, it introduces several novel aspects that were previously unexplored. In our study, we demonstrate that, in addition to laser and target parameters, the timing offset between the electron beam and the laser pulse plays a crucial role in controlling the charge coupling and acceleration processes. Thus, our findings indicate that the synchronization of timings between successive stages in a multi-stage LWFA setup significantly impacts charge coupling efficiency and the overall quality of the beam. Furthermore, we explore the evolution of beam properties, such as peak energy gain, energy spread, and beam emittance, throughout the acceleration process in the booster stage, identifying the key system parameters that affect beam quality. The study also analyzes how the initial characteristics of the electron beam influence the acceleration process and the resulting beam quality. Additionally, we investigate the effects of off-axis injection \cite{popp2010all, schnell2013optical} of the external beam relative to the laser, providing new insights into the dynamics of multi-stage LWFA.

The rest of this paper is organized as follows: Section \ref{simu} provides details of the PIC simulation setup and the simulation parameters. Section \ref{rd} and its subsections present the simulation results and discussions, focusing on the coupling and acceleration of externally injected electron beams in a laser-driven booster stage. Sections \ref{effect_target} and \ref{effect_beam} highlight the effects of plasma parameters, offset timing, and initial beam parameters on the observed phenomena. In Section \ref{effect_OA}, we discuss the impact of off-axis injection of the external beam. Finally, Section \ref{summary} provides a summary of the simulation results and concluding remarks.

\section{Particle-In-Cell Simulation Setup}\label{simu}

In this study, we performed two-dimensional (2D) Particle-In-Cell (PIC) simulations to investigate laser wakefield acceleration (LWFA) using an external injection mechanism. The open-source, fully relativistic, and massively parallelized PIC code, EPOCH \cite{arber2015contemporary, bennett2017users}, was used for this purpose.  EPOCH utilizes a second-order Yee scheme \cite{yee1966numerical} for the field solver and the standard Boris algorithm \cite{boris1970relativistic}, coupled with a modified leapfrog method \cite{arber2015contemporary}, for particle pusher.

The simulation window is a 2D rectangular box with size of $150\lambda_L \times 200\lambda_L$ in the $x$-$y$ plane, where $\lambda_L$ is the laser wavelength. A moving window scheme was used for the simulations. The grid sizes are set to $\Delta x = \lambda_L/40$ and $\Delta y = \lambda_L/10$ in the $\hat x$ and $\hat y$ directions, respectively, with an initial allocation of sixteen simulation particles per cell. The simulation time step, $\Delta t$, is determined by the Courant-Friedrichs-Lewy (CFL) condition: $C = c\Delta t /\Delta x + c\Delta t /\Delta y \leq 1$, where $C$ is the Courant number and $c$ is the speed of light in a vacuum. For this study, a Courant number of $C = 0.99$ (i.e., very close to $1.0$) was chosen to minimize numerical dispersion, resulting in a simulation time step of $\Delta t = 0.053$ fs. Open boundary conditions were applied for both electromagnetic waves and simulation particles.

\begin{figure*}
  \centering
  \includegraphics[width=4.5in,height=3.0in]{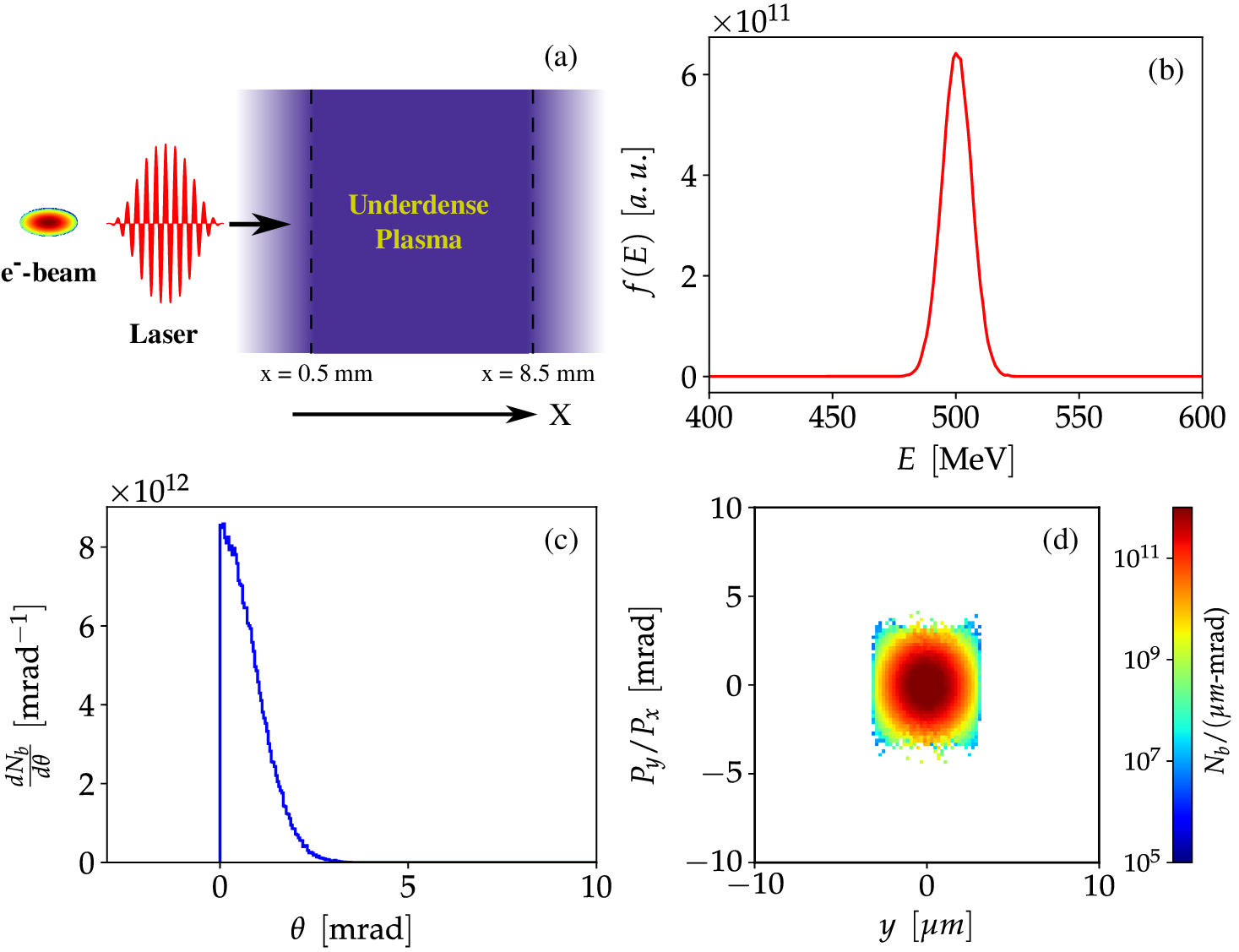}
  \caption{(a) A schematic of the PIC simulation configuration of LWFA based on the external injection mechanism. (b)-(d) Initial kinetic energy distribution, angular distribution with respect to the propagation direction (i.e., $\hat x$), and transverse phase-space diagram of the externally injected electron beam, respectively.}
\label{fig_sch_beam}
\end{figure*}

The schematic of the simulation configuration is illustrated in Fig. \ref{fig_sch_beam}(a). An intense laser pulse, followed by an electron beam, propagates into an underdense plasma medium, representing the booster stage (where the plasma frequency, $\omega_{pe}$, is lower than the laser frequency, $\omega_L$). The plasma is considered as being composed of electrons, with ions treated as a stationary, neutralizing background. The plasma target has a trapezoidal density profile along the laser propagation direction (i.e., $\hat x$) with a $8.0$-mm-long plateau and $0.5$ mm ramps. The plateau starts from $x = 0.5$ mm and extends up to $x = 8.5$ mm. Such a plasma target can be experimentally achieved using a approximately $1.5$-cm-long capillary discharge setup \cite{PhysRevResearch.6.013290}. In our simulations, the electron density of the plasma plateau is varied between $n_0 = 0.5 \times 10^{18}$ cm$^{-3}$ and $2.0 \times 10^{18}$ cm$^{-3}$ to investigate the effects of different target densities on the phenomena observed in this study. We have considered a sufficiently low and uniform plasma density to prevent the injection of unwanted electrons into the wake structure through the self-injection mechanism. The down-ramp plasma density near the exit ($x>8.5$ mm) does not contribute to electron injection, as the driver laser pulse depletes most of its energy before reaching this area.

The laser pulse is assumed to have a Gaussian profile in both the longitudinal (along the propagation direction) and transverse directions. The central wavelength of the laser pulse is $\lambda_L = 0.8$ $\mu$m, and the Full-Width-Half-Maximum (FWHM) pulse duration is $\tau_{fwhm} = 30$ fs. The laser pulse is set to propagate along $+ \hat x$ direction from the left boundary ($x = 0$) of the simulation box and is focused at the start of the plateau, i.e., at $x = 0.5$ mm. At this focal point, the laser beam has a waist size of $w_0 = 18$ $\mu$m. The peak power of the laser pulse at focus is $P_0 = 100$ TW, corresponding to a peak intensity of $I_0 \approx 2.0 \times 10^{19}$ W/cm$^2$ and a peak normalized vector potential of $a_0 = 3.0$.

An external electron beam with a Gaussian density profile in both longitudinal and transverse directions is injected from the left boundary of the simulation box. The temporal offset ($\tau_d$) of the beam relative to the laser peak is varied between $\tau_d = 85$ fs and $115$ fs to determine the optimal value. The beam's maximum density is varied between $n_b = 1.0 \times 10^{18}$ cm$^{-3}$ and $2.5 \times 10^{18}$ cm$^{-3}$ to explore the impact of beam charge density. The initial characteristics of this electron beam are illustrated in Fig. \ref{fig_sch_beam}(b)-(d). The beam has an initial peak energy of $E_{pk} = 500$ MeV, with a relative energy spread (FWHM) in between $\Delta E_{ini} = 2.5 \%$ and $15 \%$. The initial root-mean-squared (rms) transverse beam size is $\sigma_y = 1.15-5.0$ $\mu$m, and an FWHM beam duration of $\tau_b = 15$ fs (corresponding to an rms bunch length of $\sigma_x = 2.7$ $\mu$m). The initial normalized rms emittance is $\epsilon_n \approx 1.0$ mm-mrad. These initial beam parameters are considered based on the electron beam properties obtained from laser wakefield acceleration (LWFA) in a previous study \cite{maity2024parametric}. In our simulations, we varied the parameters $\tau_d$, $n_b$, $\sigma_y$, and $\Delta E_{ini}$, while keeping $E_{pk}$, $\epsilon_n$, and $\tau_b$ constant at 500 MeV, 1.0 mm-mrad, and 15 fs, respectively, to investigate the effect of initial beam parameters on the phenomena observed in our study. The detailed simulation parameters used in this study have also been provided in Table \ref{table1}.

\begin{table*}
\centering
\caption{Laser, external electron beam, and target (plasma) parameters used in this study.}
\vspace{0.5cm}
\label{table1}
\resizebox{\textwidth}{!}{%
\begin{tabular}{|c|c|c|c|}
\hline
    \makecell{Wavelength\\ ($\lambda_L$)} & \makecell{Spot size (beam waist)\\ ($w_{0}$)} & \makecell{Pulse duration (FWHM)\\ ($\tau_{fwhm}$)} & \makecell{Laser peak power ($P_0$)}\\ 
\hline
    0.8 $\mu$m & 18 $\mu$m & 30 fs & $100$ TW\\ 
\hline
\hline
    \makecell{Laser frequency\\ ($\omega_L$)} & \makecell{Laser energy\\ ($E_L$)} & \makecell{Peak value of normalized\\ vector potential ($a_0$)} & Peak intensity $I_0$ (W/cm$^2$)\\ 
\hline
    $2.35\times 10^{15}$ Hz & 3.0 J & 3.0 & $2.0\times 10^{19}$\\     
\hline
\hline
\makecell{Plasma density in plateau  ($n_{0}$)} & \makecell{Beam density \\ ($n_{b}$)} & \makecell{Initial transverse beam\\ size ($\sigma_{y}$)} & \makecell{beam duration (FWHM)\\ ($\tau_b$)}\\ 
\hline
    $(0.5-2.0)\times 10^{18}$ cm$^{-3}$ & $(1.0-2.5)\times 10^{18}$ cm$^{-3}$ & $1.15-5.0$ $\mu$m & 15 fs\\
\hline 
\hline
\makecell{Initial beam energy\\ ($E_{pk}$)} & \makecell{Initial relative\\ energy spread ($\Delta E_{ini}$)} & \makecell{Net beam charge (2D)} & \makecell{Initial normalized\\ beam emittance ($\epsilon_n$)}\\ 
\hline
    500 MeV& $2.5-15 \%$ & 1.0-10 pC/$\mu$m & 1.0 mm-mrad\\
\hline 
    
\end{tabular}}
\end{table*}

\section{Results and Discussion}\label{rd}

In the present study, we have investigated laser-plasma-based electron acceleration using the external injection mechanism mimicking multi-stage LWFA configuration. As the laser pulse propagates through an underdense plasma target, i.e., a \textit{booster stage}, plasma electrons are expelled from the region of the laser pulse by so-called ponderomotive force $\mathbf{F}_p = (mc^2/2\gamma)\nabla {\bar{a^2}}$ \cite{RevModPhys.81.1229}. Here, $a$ is the laser vector potential, and $\gamma$ represents the relativistic factor. However, plasma ions remain stationary on the electron timescale due to their significantly higher inertia. As a result, a trailing plasma wakefield (a nonlinear Langmuir wave) is formed, with a phase velocity equal to the group velocity of the laser pulse in the medium, $v_g = c\sqrt{1-(\omega_{pe}^2/\omega_L^2)}$, which decreases as the plasma density $n_0$ increases. The average strength of the wakefield $E_{LW} = (\sqrt{a_0}/2)(mc\omega_{pe}/e)$ (for $a_0>1.0$) \cite{lu2007generating}, is influenced by both the laser intensity ($a_0$) and plasma density ($n_0$), and it increases with higher values of both. Conversely, the characteristic size of the wakefield, i.e., blowout radius $R = 2c\sqrt{a_0}/\omega_{pe}$, decreases with increasing $n_0$ but increases with $a_0$. An electron beam injected into this laser-driven wakefield with a sufficiently high initial velocity to match the phase velocity of the wakefield will either be accelerated or decelerated depending on the local phase of the wake electric field. Specifically, negatively charged electrons will be accelerated in the negative half-cycles and decelerated in the positive half-cycles of the electric field. Thus, the charge coupling efficiency and subsequent acceleration of an external electron beam trailing the laser pulse are strongly influenced by the plasma density ($n_0$) of the \textit{booster stage}, the initial temporal offset relative to the laser pulse ($\tau_d$), and the laser intensity ($a_0$). Furthermore, the characteristics of the electron beam will evolve as the laser intensity, spot size, and consequently the wakefield self-consistently evolve during propagation in the \textit{booster stage}. In the following subsections, we present our simulation results that highlight these phenomena. Additionally, we demonstrate the impact of the beam's initial characteristics on the coupling and acceleration process within the \textit{booster stage}. 


\subsection{Effect of Target Density and Initial Temporal Offset}\label{effect_target}

\begin{figure*}
  \centering
  \includegraphics[width=5.0in,height=2.5in]{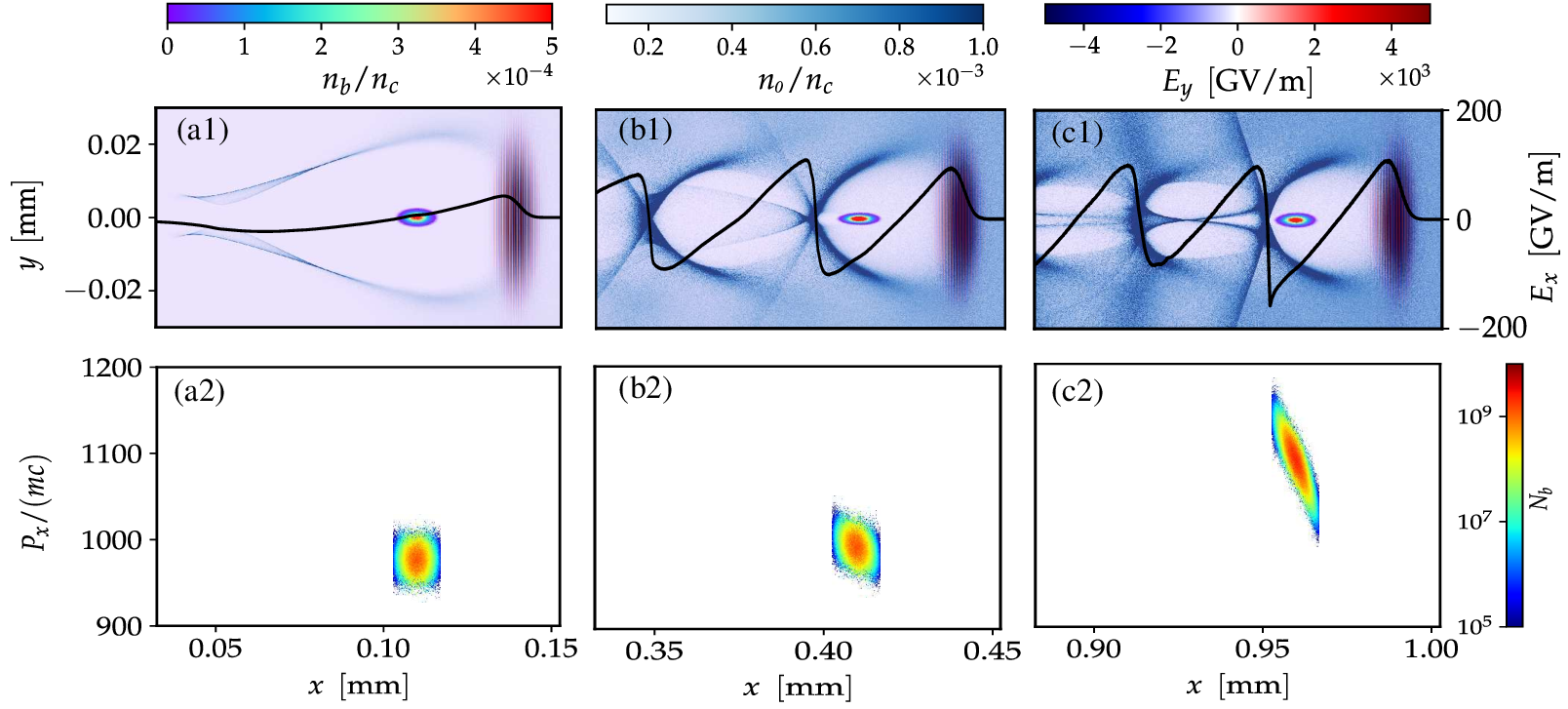}
  \caption{(a1)-(c1) Density of the external electron beam ($n_b$), plasma electron density ($n_0$), laser electric field ($E_y$), and on-axis wakefield ($E_x$) at different stages of the laser propagation through the plasma target. (a2)-(c2) The evolution of longitudinal phase-space ($P_x$-$x$) distribution of external electron beam at the corresponding propagation distances. In this case, we have initially considered the temporal offset between laser and external electron beam $\tau_d = 98$ fs (corresponding space difference, $\Delta x \approx 29.4$ $\mu$m), plasma electron density $n_{0} = 1.0\times 10^{18}$ cm$^{-3}$, transverse beam size $\sigma_y = 1.15$ $\mu$m, initial peak energy of the beam $E_{pk} = 0.5$ GeV, and initial relative energy spread $\Delta E_{ini} = 2.5\%$.}
\label{fig_ini_couplng}
\end{figure*}

\begin{figure*}
  \centering
  \includegraphics[width=5.5in,height=3.5in]{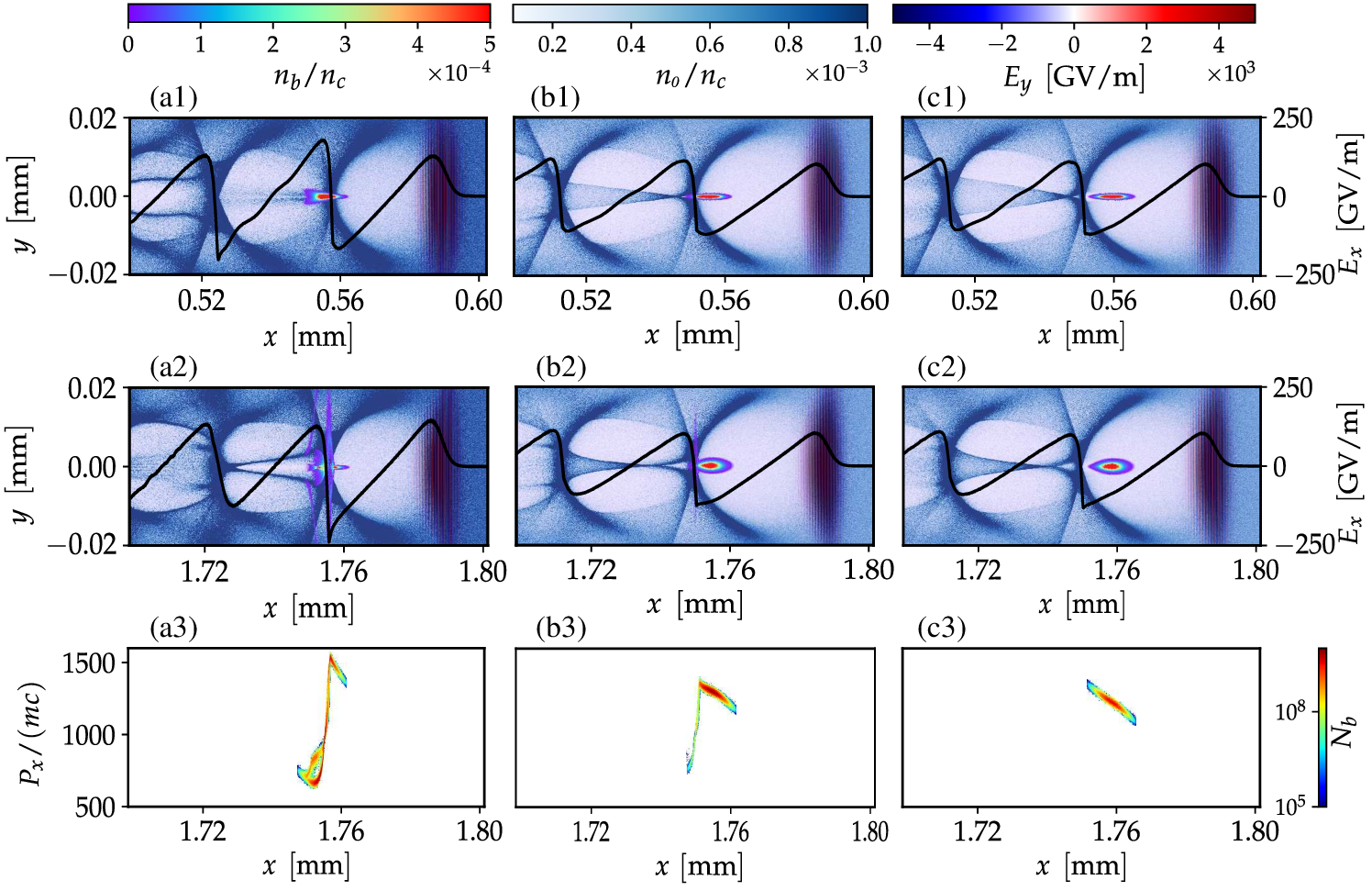}
  \caption{Density distributions of  plasma electrons ($n_0$) as well as externally injected electrons ($n_b$) in 2D x-y plane and on-axis wakefield ($E_x$) have been shown for :  (a1)-(a2) $n_0 = 1.5\times 10^{18}$ cm$^{-3}$, $\tau_d = 111.5$ fs ($\Delta x \approx 33.4$ $\mu$m); (b1)-(b2) $n_0 = 1.0\times 10^{18}$ cm$^{-3}$, $\tau_d = 111.5$ fs ($\Delta x \approx 33.4$ $\mu$m); (c1)-(c2) $n_0 = 1.0\times 10^{18}$ cm$^{-3}$, $\tau_d = 98$ fs ($\Delta x \approx 29.4$ $\mu$m) at two different instants of time $t = 2.0$ ps and $6.0$ ps, respectively. All other system parameters are kept fixed. In (a3), (b3), and (c3), longitudinal phase-space distributions of externally injected electrons have been shown at time $t = 6.0$ ps for these cases, respectively.}
\label{fig_n0td}
\end{figure*}

\begin{figure*}
  \centering
  \includegraphics[width=5.5in,height=4.0in]{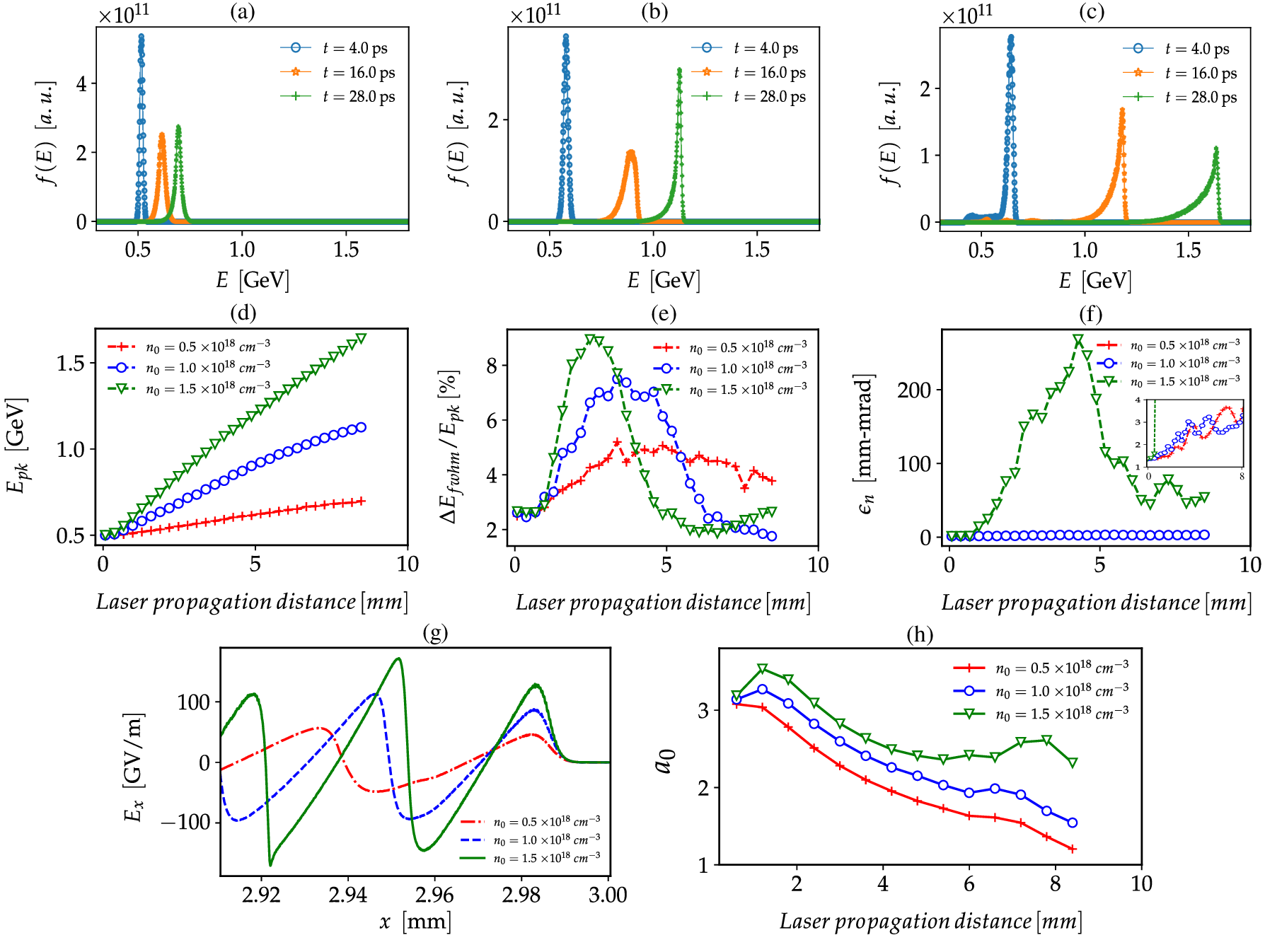}
  \caption{(a), (b), and (c) Kinetic energy distribution of externally injected electron beam at different times for simulations with $n_0 = 0.5\times 10^{18}$ cm$^{-3}$, $1.0\times 10^{18}$ cm$^{-3}$, and $1.5\times 10^{18}$ cm$^{-3}$, respectively. (d), (e), and (f) Time evolution of peak energy ($E_{pk}$), relative energy spread ($\Delta E_{fwhm}/E_{pk}$), and normalized emittance ($\epsilon_n$) of the externally injected electron beam, respectively. (g) On-axis longitudinal electric field ($E_x$) at $t = 10$ ps for these three different values of plasma density. (h) The peak value of normalized vector potential ($a_0$) as a function of laser propagation distance in these three cases. In all cases, we have used fixed initial parameters: $\tau_d = 98$ fs, $\sigma_y = 1.15$ $\mu$m, $\sigma_x = 2.7$ $\mu$m, $E_{pk} = 0.5$ GeV, $\Delta E_{ini} \approx 2.5\%$, and an initial normalized beam emittance of $\epsilon_n \approx 1.0$ mm-mrad.}
\label{fig_diag_n0}
\end{figure*}

Coupling and acceleration of an external electron beam in the laser-driven wakefield at different stages of the laser propagation through the plasma target (\textit{booster stage}) have been illustrated in Fig. \ref{fig_ini_couplng}. It is seen from Fig. \ref{fig_ini_couplng}(a1)-(c1) that ``horse-shoe-shaped" wake waves are created behind the laser pulse as it propagates through the plasma. This is a consequence of increase in nonlinear plasma wavelength with increasing laser intensity which peaks on the laser propagation axis and has a Gaussian radial profile. It is also seen that the external electron beam is injected into the accelerating phase (negative half-cycles) of the first wake wave behind the laser pulse. The results presented in Fig. \ref{fig_ini_couplng} are obtained from a specific simulation run using optimal parameters. The external electron beam has a transverse size of $\sigma_y = 1.15$ $\mu$m, while the transverse size (blowout radius) of the wake wave, approximately equal to the laser spot size, is $R \approx w_0 = 18$ $\mu$m. This allows the entire electron beam to fit within a single wake structure. At the plateau, the plasma density is $n_0 = 1.0 \times 10^{18}$ cm$^{-3}$, corresponding to a linear plasma wavelength of $\lambda_p \approx 33.4$ $\mu$m. The temporal offset of the beam relative to the laser is $\tau_d = 98$ fs, corresponding to an initial laser-beam separation of $\Delta x \approx 29.4$ $\mu$m. As will be discussed later, this value of $\tau_d$, along with the plasma density $n_0$, is identified as the optimal parameters for injecting the entire electron beam within the accelerating phase of the wakefield. The external electron beam has an longitudinal rms length of $\sigma_x = 2.7$ $\mu$m, ensuring that all electrons are injected into the accelerating phase of the wakefield. The longitudinal phase-space ($P_x$-$x$) evolution of the beam with the laser propagation distance has been depicted in Fig. \ref{fig_ini_couplng}(a2)-(c2). Initially, up to a propagation distance of $x = 0.5$ mm, the injected electron beam acquires negligible energy, as illustrated in Fig. \ref{fig_ini_couplng}(a2) and (b2). This is due to the lower plasma density in this region, which results in a weaker wakefield. However, once the beam enters the plateau region ($x > 0.5$ mm) where the plasma density is higher, it gains over 50 MeV of energy within just 0.5 mm of propagation, as demonstrated in Fig. \ref{fig_ini_couplng}(c2).

As pointed out in the above discussion, the charge coupling efficiency is strongly influenced by the plasma density ($n_0$) in the booster stage and the initial temporal offset ($\tau_d$) of the beam with respect to the peak of the laser pulse. This effect is explicitly illustrated in Fig. \ref{fig_n0td}. We define a dimensionless parameter $\Lambda = [(\lambda_p-c\tau_d) - \sigma_x]/\lambda_L$. This parameter represents the relative position of the injected electron beam within the first bubble (wake) formed behind the laser driver. For effective injection of the entire electron beam into the accelerating (negative-half) phase of the wakefield, the condition $\Lambda>0$ must be satisfied. First, we present the results of a simulation with $n_0 = 1.5 \times 10^{18}$ cm$^{-3}$ and $\tau_d = 111.5$ fs analyzed at two different times $t = 2.0$ ps and $t = 6.0$ ps , as shown in Fig. \ref{fig_n0td}(a1) and (a2), respectively. In this case, $\Lambda \approx -11.1$, indicating a negative value. Consequently, as seen in Fig. \ref{fig_n0td}(a1), a significant portion of the external beam is injected not into the first bubble but into the decelerating and defocusing phase of the second bubble. As a result, at later times, many of the injected electrons become unfocused and scatter in the transverse direction, as illustrated in Fig. \ref{fig_n0td}(a2). Additionally, these electrons experience deceleration, which is evident from the longitudinal phase-space distribution presented in Fig. \ref{fig_n0td}(a3). We then reduced the plateau density of the plasma to $n_0 = 1.0 \times 10^{18}$ cm$^{-3}$, while keeping all other initial parameters unchanged, e.g., $\tau_d = 111.5$ fs. The simulation results for this scenario are presented in Fig. \ref{fig_n0td}(b1)-(b3). It can be observed that most electrons in the external beam are now injected into the accelerating phase of the wake wave, resulting in momentum gain. This occurs because decreasing the plasma density increases the size of the wake (i.e., $\lambda_p$). In this case, although the negative value of $\Lambda$ is reduced ($\Lambda \approx -3.4$), it still does not satisfy the condition $\Lambda>0$. As a result, as shown in Fig. \ref{fig_n0td}(b2) and (b3), some electrons from the external beam are still injected into the decelerating and defocusing phase of the wakefield, leading to transverse scattering and deceleration. Next, we conducted a simulation with a reduced temporal offset of $\tau_d = 98$ fs between the beam and the laser, along with a lower plasma density of $n_0 = 1.0 \times 10^{18}$ cm$^{-3}$. The results are shown in Fig. \ref{fig_n0td}(c1)-(c3). In this case, $\Lambda \approx 1.6$, satisfying the condition $\Lambda>0$. Consequently, the entire electron beam is effectively injected into the accelerating phase of the wakefield and undergoes acceleration. It is important to note that the parameter $\Lambda$ is obtained from the linear analysis, where we assume that the wake size is on the order of the plasma wavelength $\lambda_p$. In highly nonlinear scenarios, such as the bubble regime, this parameter also becomes dependent on $a_0$. This dependency arises because, in the nonlinear regime, the wake size is also influenced by $a_0$, increasing with higher values of this parameter.  It positively affects the charge coupling process since, in the nonlinear stage, the bubble radius (dependent on $a_0$) exceeds the plasma wavelength $\lambda_p$ \cite{lu2007generating}. Therefore, the dependence of $\Lambda$ on $a_0$ does not change the charge coupling mechanism but enhances its efficiency. However, $a_0$ can still influence the acceleration process and beam quality through the beam loading effect, with both the amplitude and slope of the accelerating field being dependent on $a_0$ in the nonlinear regime. In conclusion, in experiments, one can adjust plasma density, laser intensity, and/or the relative temporal offset between the laser and the external beam to control the efficient injection of the beam into the wakefield.

We have analyzed the evolution of the externally injected electron beam's properties as it propagates through the plasma medium in the wake of the laser pulse. Figure \ref{fig_diag_n0} illustrates the time history of the external electron beam’s characteristics for various cases, with different plasma densities ($n_0$) in the booster stage. 

Figure \ref{fig_diag_n0}(a)-(c) illustrates the energy distribution of the beam at three different stages of its evolution for plasma densities of $n_0 = 0.5 \times 10^{18}$ cm$^{-3}$, $1.0 \times 10^{18}$ cm$^{-3}$, and $1.5 \times 10^{18}$ cm$^{-3}$, respectively. All other system parameters were kept constant: $\tau_d = 98$ fs, $\sigma_y = 1.15$ $\mu$m, $\sigma_x = 2.7$ $\mu$m, $E_{pk} = 0.5$ GeV, and $\Delta E_{ini} = 2.5\%$. In these cases, the parameter $\Lambda$ is calculated to be $\Lambda \approx 19$, $1.6$, and $-6$, respectively. In all cases, it is observed that the peak energy of the externally injected electron beam increases over time, i.e., with increasing propagation distance. The peak energy gain is significantly higher at higher plasma densities. This is attributed to the fact that the strength of the wakefield driven by the laser pulse intensifies as plasma density increases. However, for high plasma densities, e.g., $n_0 = 1.5\times 10^{18}$ cm$^{-3}$ ($\Lambda<0$), a significant portion of the beam fails to be injected into the accelerating phase of the wakefield, resulting in deceleration and particle loss, as illustrated in Fig. \ref{fig_diag_n0}(c). This effect also contributes to the broadening of the energy distribution, leading to a higher absolute energy spread over time. This observation is consistent with the phenomena discussed in Fig. \ref{fig_n0td}.

The peak energy ($E_{pk}$) of the beam is tracked as it increases with propagation distance and is explicitly illustrated in Fig. \ref{fig_diag_n0}(d) for different values of $n_0$. The rate of increase in $E_{pk}$ over the laser propagation distance is significantly higher at greater plasma densities. Consequently, the final peak energy gained by the electron beam over a fixed propagation distance is substantially larger for higher plasma densities. As mentioned earlier, this effect is due to the increased wakefield strength at higher plasma densities, as illustrated in Fig. \ref{fig_diag_n0}(g). 

The relative energy spread ($\Delta E_r$) also evolves differently depending on plasma density, as shown in Fig. \ref{fig_diag_n0}(e). The relative energy spread of the beam is defined as $\Delta E_r = \Delta E_{fwhm}/E_{pk}$, where $\Delta E_{fwhm}$ represents the absolute energy spread measured at the full width at half maximum (FWHM) of the beam’s energy distribution. The relative energy spread was initially set at $\Delta E_r \approx 2.5\%$ for all cases. In each case, $\Delta E_r$ initially increases from this value to a maximum, then decreases as the beam propagates through the plasma target. Notably, the maximum value of $\Delta E_r$ is higher and is reached earlier at higher plasma densities. The evolution of $\Delta E_r$ with propagation distance can be explained as follows. The value of $\Delta E_r$ ($\Delta E_{fwhm}/E_{pk}$) changes as both $\Delta E_{fwhm}$ and $E_{pk}$ vary independently. Since $E_{pk}$ increases almost linearly with the propagation distance, the nature of evolution of $\Delta E_r$ is primarily determined by the absolute energy spread $\Delta E_{fwhm}$. As soon as the external electron beam is injected into the wake, the absolute energy spread $\Delta E_{fwhm}$ starts to increase. This is because the longitudinal electric field $E_x$ within the wake structure is non-uniform and exhibits a linear slope. Consequently, electrons within the beam, located at different positions inside the wake structure, experience varying accelerating fields, leading to an increase in energy spread across the beam. The on-axis longitudinal wake electric field profiles at any particular instant of time $t = 10$ ps are depicted in Fig. \ref{fig_diag_n0}(g) for different cases with varying plasma plateau densities. It is observed that the slope of $E_x$ is steeper for higher plasma densities. As a result, $\Delta E_{fwhm}$ increases more rapidly with propagation distance, reaching a higher maximum value for the cases with higher plasma density. After a certain propagation distance, $\Delta E_{fwhm}$ begins to decrease from its maximum value. This decrease is attributed to the electric field originated from the beam’s space-charge, which modifies the slope of the laser-driven wakefield, rendering it more uniform across the beam length \cite{PhysRevLett.103.194804}. Additionally, the decrease in the normalized vector potential $a_0$ with propagation distance, as depicted in Fig. \ref{fig_diag_n0}(h), reduces the strength of the wake electric field, resulting in a decrease in the absolute energy spread. Ultimately, $\Delta E_{fwhm}$ approaches saturation at higher plasma densities after sufficient propagation. This saturation occurs due to the stabilization of the laser spot size (after $x \sim 6.0$ mm, as shown in Fig. \ref{fig_diag_n0}(h)) as a result of the self-consistently developed self-focusing effect \cite{sprangle1987relativistic, PhysRevA.41.4463}, which leads to a stable slope of $E_x$. The final values of $\Delta E_r$ are lower for higher plasma densities because of the larger values of $E_{pk}$.

Another important parameter in particle beam analysis is the normalized transverse trace-space emittance, which characterizes the transverse beam quality. It is defined as \cite{PhysRevSTAB.16.011302} $\epsilon_n = \sqrt{\langle y^2 \rangle \langle \beta^2 \gamma^2 y'^2 \rangle - {\langle y \beta \gamma y' \rangle}^2}$, where $\beta = v/c \approx 1.0$ for a relativistic beam, with $v$ representing the particle velocity and $c$ the speed of light in vacuum. In this expression, $\gamma$ is the relativistic factor, while $y$ and $y'$ represent the particle's transverse position and its angular divergence with respect to the beam axis, respectively. Figure \ref{fig_diag_n0}(f) shows the evolution of beam emittance as a function of propagation distance for three different cases with varying plasma densities. In all cases, an initial normalized beam emittance of $\epsilon_n \approx 1.0$ mm-mrad was assumed. For plasma densities of $n_0 = 0.5 \times 10^{18}$ cm$^{-3}$ and $1.0 \times 10^{18}$ cm$^{-3}$, the beam emittance remains nearly constant throughout the evolution, increasing only slightly from $\epsilon_n \approx 1.0$ mm-mrad to 3.0 mm-mrad over a propagation distance of 8 mm, as shown in the inset of Fig. \ref{fig_diag_n0}(f). Notably, the emittance growth trend is similar in both cases, indicating that it is independent of the plasma density ($n_0$). However, in the case of $n_0 = 1.5 \times 10^{18}$ cm$^{-3}$, the normalized beam emittance increases significantly with propagation distance. This dramatic growth is attributed to the fact that not all electrons in the beam are injected into the accelerating phase of the first plasma bubble. A substantial fraction of the external electron beam is injected into the decelerating and defocusing phase (the electron sheath of the first bubble), leading to increased emittance. Therefore, as long as all electrons of the beam are injected into the accelerating and focusing phase of the wakefield, the normalized beam emittance remains independent of the plasma density.

The above analysis highlights the typical trade-offs between the phase and amplitude of the wake electric field in optimizing electron acceleration in LWFA setups. For efficient acceleration, the entire electron beam must remain within the accelerating phase of the wakefield behind the laser pulse. Being too close to the laser pulse can cause deceleration due to phase slippage, particularly at higher energies, while being too far back risks falling out of the wakefield or encountering a defocusing field. Ideally, electrons should be aligned at the optimal phase, near the negative peak of the accelerating field, to maximize energy gain and minimize energy spread. A stronger electric field (at higher plasma densities) accelerates electrons more rapidly, allowing higher energy gains over shorter distances. However, increasing the wakefield's amplitude can alter the phase velocity of the wake, causing electrons to phase slip to the decelerating phase within a short distance. A higher wakefield amplitude also increases energy spread due to a steeper slope. High amplitude may also lead to nonlinear effects such as beam loading and wave breaking, increasing energy spread, and trapping unwanted electrons. Therefore, carefully tuning plasma density, laser intensity, and injection timing is essential for achieving efficient acceleration and maintaining the quality of the electron beam.


\subsection{Effect of Initial Beam Parameters}\label{effect_beam}

\begin{figure*}
  \centering
  \includegraphics[width=5.5in,height=4.0in]{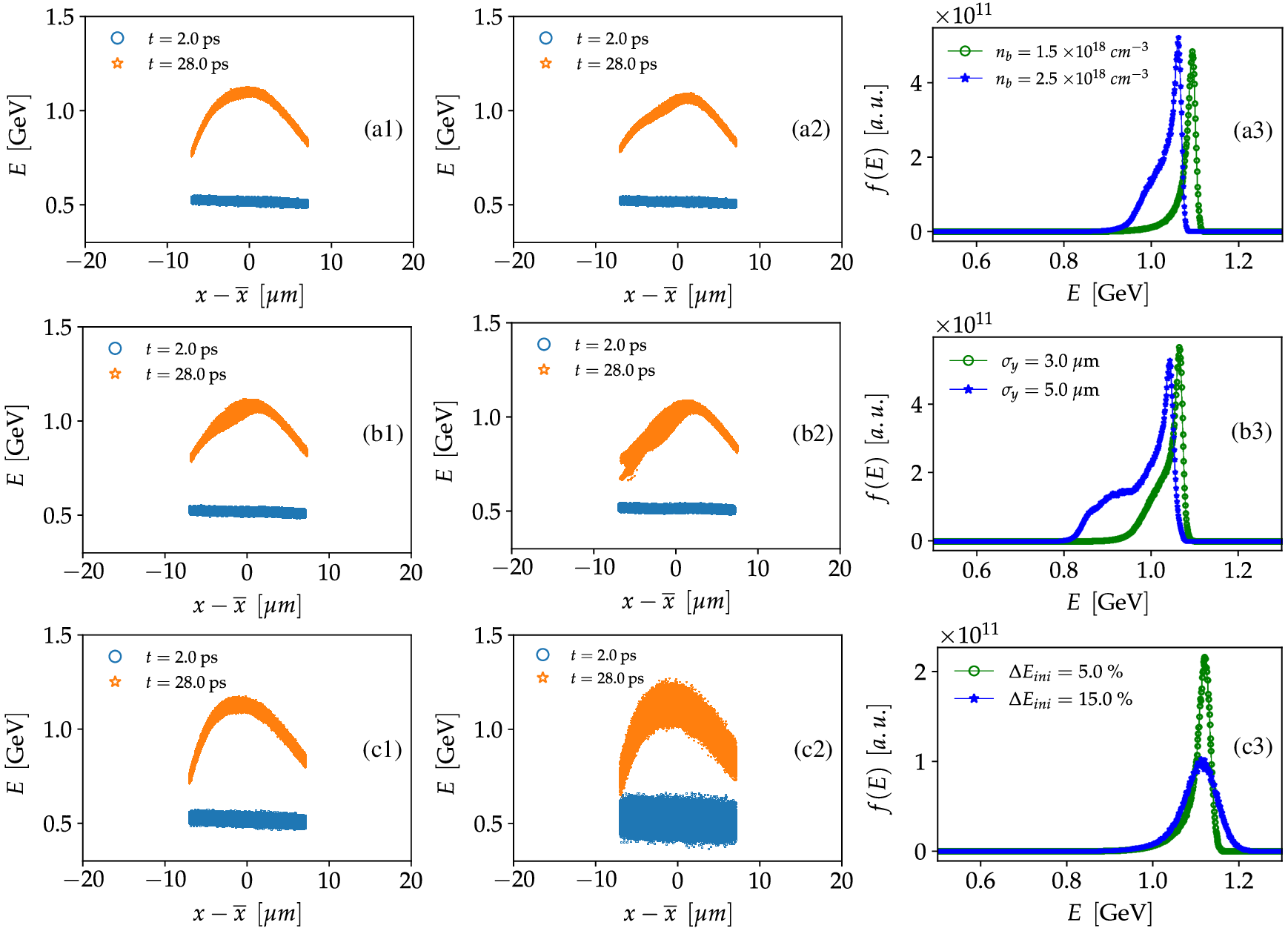}
  \caption{Particle distribution in E-x plane with two different values of initial beam density (a1) $n_b = 1.5\times 10^{18}$ cm$^{-3}$ and (a2) $n_b = 2.5\times 10^{18}$ cm$^{-3}$, keeping all other initial parameters fixed, i.e., initial transverse size of the beam $\sigma_y = 1.15$ $\mu$m, plasma density $n_p = 1.0\times 10^{18}$ cm$^{-3}$, and initial energy spread of the beam $\Delta E_{ini} = 2.5 \%$. In (a3), the energy histograms of the particles at the final time ($t = 28$ ps) of these simulation runs have been shown. The same has been depicted in subplots (b1)-(b3) with two different initial values of transverse beam size $\sigma_y = 3.0$ $\mu$m and $5.0$ $\mu$m, respectively, with fixed $n_b = 1.0\times 10^{18}$ cm$^{-3}$, $n_p = 1.0\times 10^{18}$ cm$^{-3}$, and $\Delta E_{ini} = 2.5 \%$. Similarly, subplots (c1)-(c3) demonstrate the same for two different values of initial energy spread of the beam $\Delta E_{ini} = 5.0 \%$ and $15.0 \%$, respectively, for fixed $n_b = 1.0\times 10^{18}$ cm$^{-3}$, $n_p = 1.0\times 10^{18}$ cm$^{-3}$, and $\sigma_y = 1.15$ $\mu$m. }
\label{fig_eng_x_beam}
\end{figure*}

\begin{figure*}
  \centering
  \includegraphics[width=5.5in,height=3.0in]{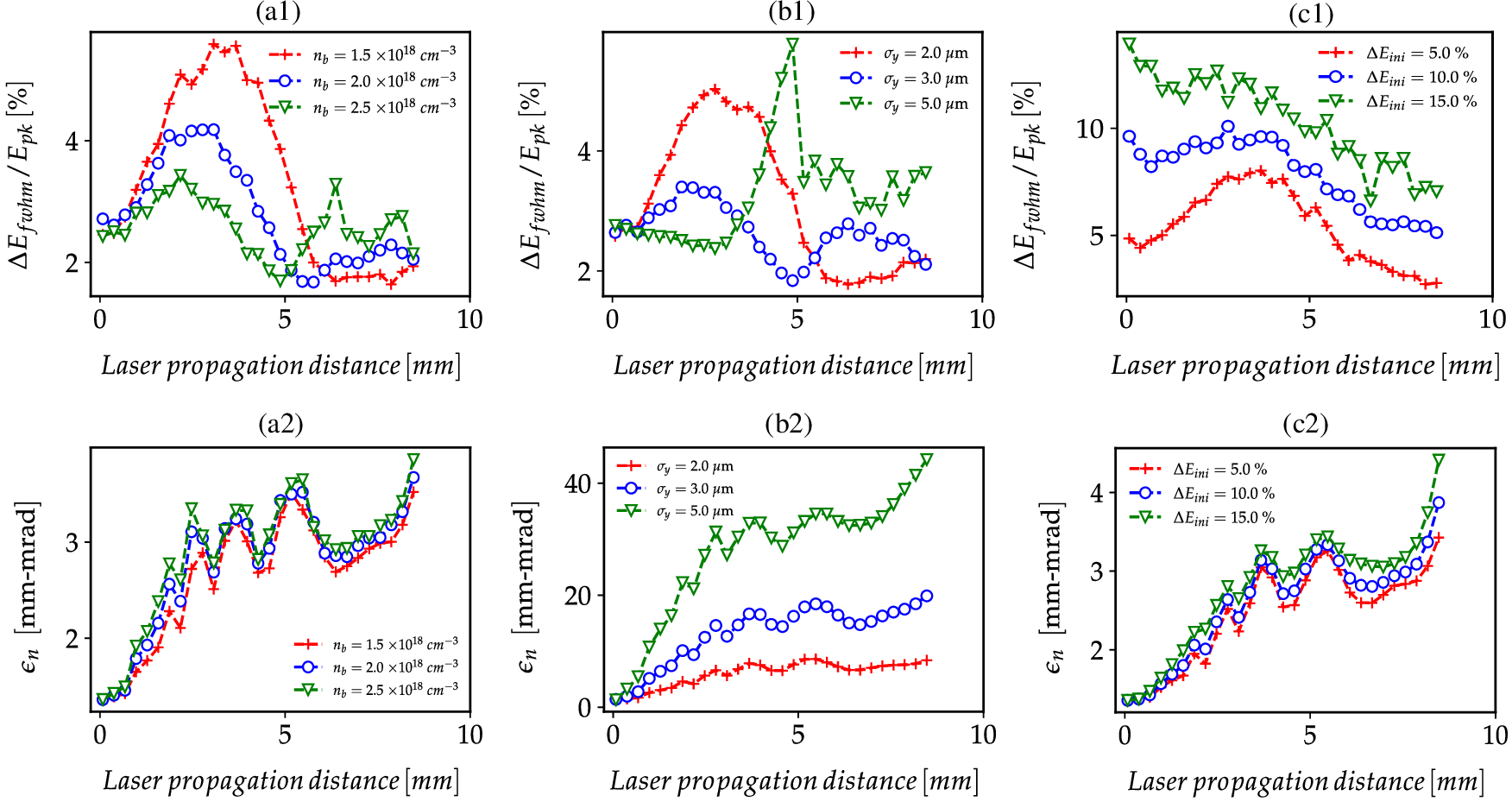}
  \caption{ Evolution of externally injected electron beam properties, i.e., FWHM relative energy spread ($\Delta_{fwhm}/E_{pk}$), and normalized beam emittance ($\epsilon_n$) as a function of laser propagation distance for different initial values of beam parameters, i.e., $n_b$, $\sigma_y$, and $\Delta E_{ini}$.}
\label{fig_diag_beam}
\end{figure*}

The initial characteristics of the beam may influence the charge coupling, the acceleration process, and the overall quality of the accelerated beam. This effect has been thoroughly investigated, and the results are presented in Fig. \ref{fig_eng_x_beam} and Fig. \ref{fig_diag_beam}.

The longitudinal energy-space ($E$-$x$) distribution of electrons in the externally injected beam is shown for two different beam densities: $n_b = 1.5 \times 10^{18}$ cm$^{-3}$ and $n_b = 2.5 \times 10^{18}$ cm$^{-3}$, in Fig. \ref{fig_eng_x_beam}(a1) and (a2), respectively. All other simulation parameters were kept constant. In both cases, it was observed that all the electrons in the beam effectively injected into the wake electric field and gained acceleration. Thus, the charge coupling is not influenced by the initial beam density. The final energy distribution function of the beam, at the end of the plateau at $t = 28$ ps, is depicted in Fig. \ref{fig_eng_x_beam}(a3) for these two cases. For higher beam densities, the final peak energy achieved by the beam is slightly lower, and the energy distribution is broader, indicating a higher energy spread. This result is attributed to the space-charge effects \cite{PhysRevLett.103.194804} of the externally injected electron beam. The longitudinal component of the self-induced electric field of the beam, which is proportional to the beam's charge, influences both the strength and slope of the wake electric field. As a result, both the peak energy gained by the beam and the energy spread are affected by the beam density. A similar trend is shown in Fig. \ref{fig_eng_x_beam}(b1)-(b3) for varying transverse beam sizes ($\sigma_y$), where increasing $\sigma_y$ leads to higher beam charge. Consequently, as $\sigma_y$ increases, the peak energy decreases, and the energy spectrum broadens. Thus, although charge coupling and subsequent acceleration do not depend on the initial value of $\sigma_y$, the final energy gained and the energy spread are influenced by it. Additionally, our study indicates that the initial energy spread ($\Delta E_{ini}$) of the externally injected electron beam has no impact on charge coupling, acceleration, or the final energy distribution of the beam, as illustrated in Fig. \ref{fig_eng_x_beam}(c1)-(c3).

We have also tracked the evolution of the two important characteristics of the beam, i.e., relative energy spread $\Delta E_r$ and normalized emittance $\epsilon_n$ as a function of propagation distance. The results, presented in Fig. \ref{fig_diag_beam}, correspond to different cases with varying initial beam parameters, such as $n_b$, $\sigma_y$, and $\Delta E_{ini}$. Since the evolution of peak energy remains largely unaffected by $n_b$, $\sigma_y$, and $\Delta E_{ini}$, the variation in relative energy spread ($\Delta E_r$), shown in Fig. \ref{fig_diag_beam}(a1)-(c1), is primarily determined by the absolute energy spread ($\Delta E_{fwhm}$). For the case with varying $n_b$ (Fig. \ref{fig_diag_beam}(a1)), the absolute energy spread ($\Delta E_{fwhm}$) initially increases with propagation distance, reaches a maximum, and then decreases. The maximum value of $\Delta E_{fwhm}$ is higher for lower $n_b$ values. This occurs because, at higher $n_b$ (corresponding to higher beam charge), the slope of the longitudinal wakefield becomes less steep due to the space-charge electric field of the beam. Over a sufficient propagation distance, the self-consistent interaction between the beam density, space-charge field, and wakefield causes $\Delta E_{fwhm}$ to decrease and eventually stabilize. Ultimately, the final values of $\Delta E_{fwhm}$ for different $n_b$ values are nearly identical, with minor variations attributed to small differences in peak energy ($E_{pk}$). A similar trend is observed in Fig. \ref{fig_diag_beam}(b1) for cases with varying $\sigma_y$. However, for very large transverse beam sizes (e.g., $\sigma_y = 5.0$ $\mu$m), there is a noticeable increase in $\Delta E_{fwhm}$ after a certain propagation distance ($x\approx 5.0$ mm). The exact reason for this behavior is not fully understood. It may be related to the fact that the longitudinal wake electric field is not uniform but depends on the bubble radius $r$. For sufficiently large transverse beam sizes, the impact of this non-uniformity on beam quality becomes more significant. Figure \ref{fig_diag_beam}(c1) illustrates the variation of $\Delta E_r$ with propagation distance for different initial relative energy spreads $\Delta E_{ini}$. For very high values of $\Delta E_{ini}$, such as $10 \%$ and $15\%$, the relative energy spread $\Delta E_r$ decreases with propagation distance due to the increase in $E_{pk}$. For lower values of $\Delta E_{ini}$, $\Delta E_r$ initially increases before decreasing. This behavior is consistent with the results shown in Fig. \ref{fig_diag_n0}(e). We have also performed several PIC simulations using different initial values of beam emittance while keeping all other parameters constant. The simulation results (not shown here) indicate no correlation between initial beam emittance and the charge coupling efficiency or resultant beam quality.

The normalized beam emittance as a function of propagation distance is shown in Fig. \ref{fig_diag_beam}(a2)-(c2). It is evident that the evolution of $\epsilon_n$ is largely independent of both $n_b$ and $\Delta E_{ini}$, as illustrated in Fig. \ref{fig_diag_beam}(a2) and (c2). For all cases with varying $n_b$ and $\Delta E_{ini}$, $\epsilon_n$ increases slightly from 1.0 mm-mrad to approximately 4.0 mm-mrad. This growth is attributed to a slight mismatch between the beta function $\beta_c =\sigma_y^2 /\epsilon$ of the beam relative to the betatron wavelength ($\lambda_{\beta}$) \cite{PhysRevSTAB.15.111303}. Here, $\epsilon = \sqrt{\langle y^2 \rangle \langle y'^2 \rangle - {\langle yy' \rangle}^2} \approx \epsilon_n/\langle \gamma \rangle$ represents the transverse trace-space emittance. During acceleration, individual electrons in the beam undergo transverse betatron oscillations with a frequency $\omega_{\beta} \approx \omega_{pe}/\sqrt{2\gamma}$ and a corresponding wavelength $\lambda_{\beta} = 2\pi c/\omega_{\beta}$ \cite{liang2023characteristics}. The oscillations observed in emittance across all cases are a manifestation of the betatron motion of individual particles, with a wavelength closely matching the theoretically calculated betatron wavelength for the given system parameters, i.e., $\lambda_{\beta} \approx 1.4$ mm. This betatron motion will lead to emittance growth through a process known as betatron-phase mixing \cite{PhysRevSTAB.15.111303}. Such emittance growth can be mitigated by using a matched beta function $\beta_m \approx \lambda_{\beta}$ of the beam. In our case, for the chosen system parameters, the matched transverse beam size is $\sigma_{ym} \approx 1.2$ $\mu$m. Therefore, as $\sigma_y$ is increased, moving further away from this matching condition, the growth of $\epsilon_n$ becomes more prominent, as shown in Fig. \ref{fig_diag_beam}(b2).


\subsection{Effect of Off-Axis Injection}\label{effect_OA}

\begin{figure*}
  \centering
  \includegraphics[width=4.5in,height=3.0in]{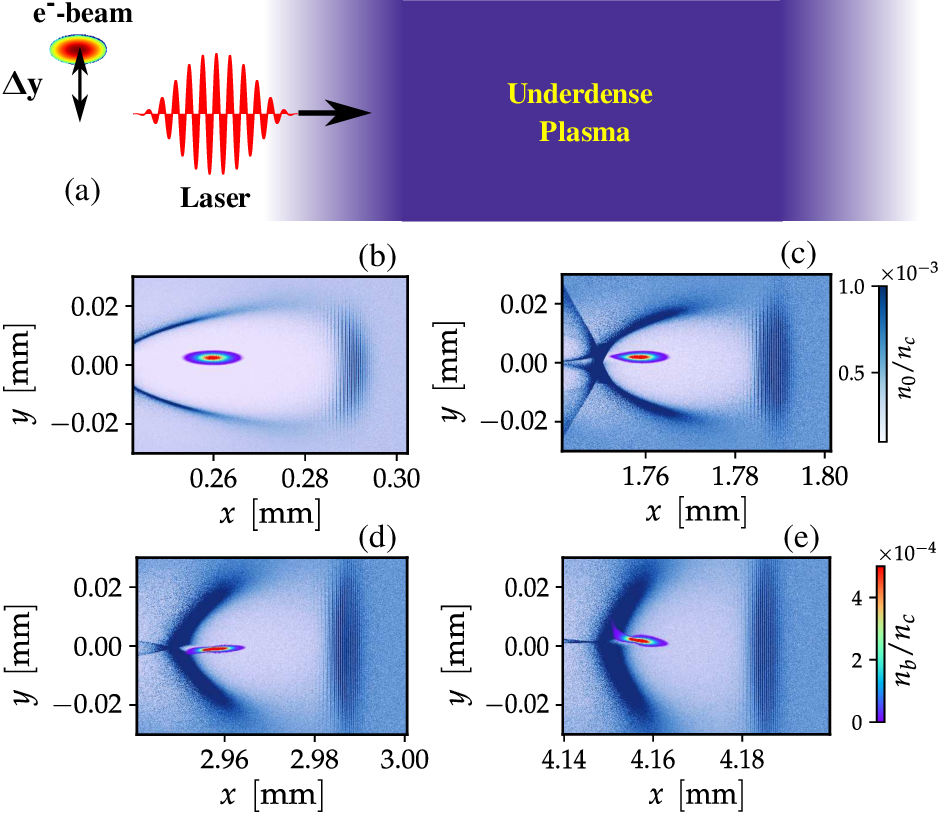}
  \caption{(a) A schematic of the simulation set up for off-axis injection of external electron beam. (b)-(e) Electron density of the target plasma and external beam in 2D x-y plane at four different times of evolution. In this particular case, the transverse distance ($\Delta y$) between the axis of the laser propagation ($y = 0$) and axis of the external e-beam is considered to be $\Delta y = 3.0$ $\mu$m with fixed $n_b = 1.0\times 10^{18}$ cm$^{-3}$, $n_p = 1.0\times 10^{18}$ cm$^{-3}$, $\tau_d = 98$ fs, $\sigma_y = 1.15$ $\mu$m, $E_{pk} = 0.5$ GeV, and $\Delta E_{ini} = 2.5 \%$.}
\label{fig_OA_sch}
\end{figure*}

\begin{figure*}
  \centering
  \includegraphics[width=5.0in,height=3.0in]{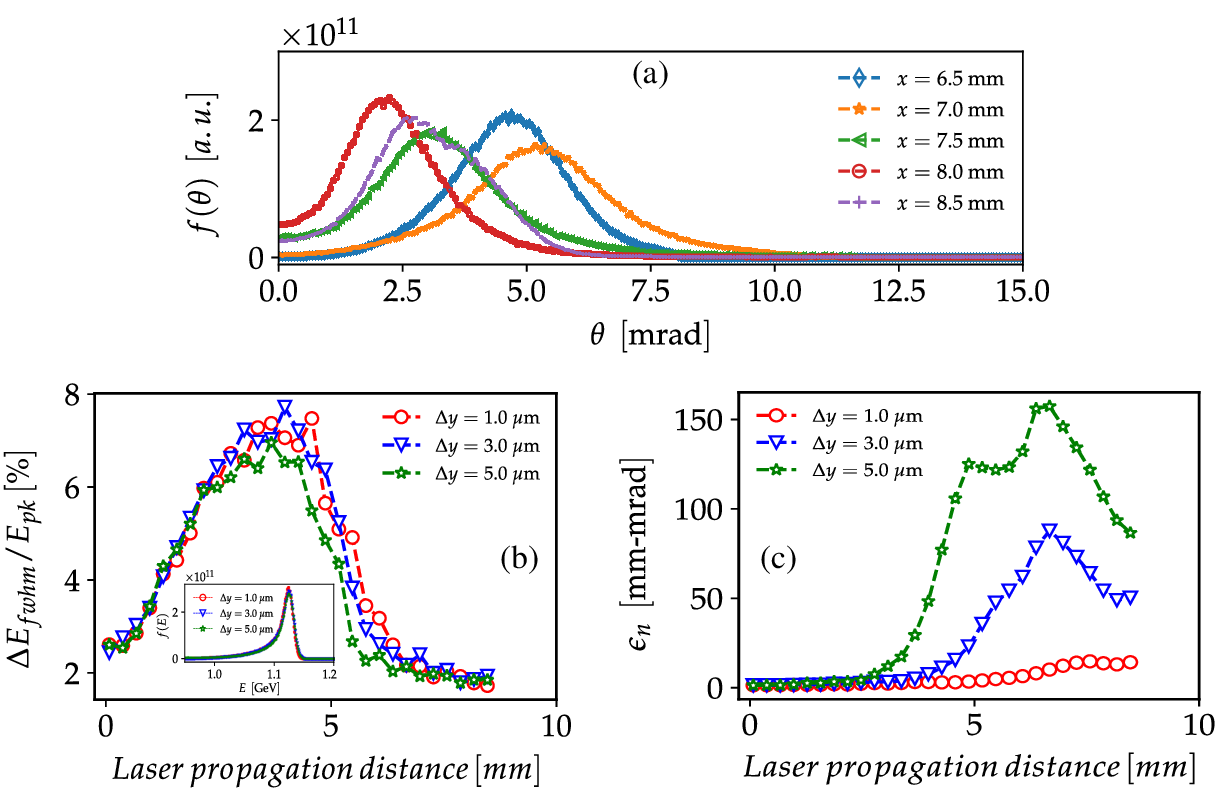}
  \caption{(a) The angular distributions of externally injected electron beam at different laser propagation distance ($x$) for a particular simulation with $\Delta y = 3.0$ $\mu$m. (b) Evolution of relative energy spread ($\Delta E_{fwhm}/E_{pk}$), and (c) normalized emittance ($\epsilon_n$) as a function of laser propagation distance for different simulation runs with changing values of initial off-axis distance $\Delta y$. The inset in (b) shows energy histogram of the externally injected electrons at final time ($t = 28$ ps) of the evolution. In these simulation runs, we have considered fixed $n_b = 1.0\times 10^{18}$ cm$^{-3}$, $n_p = 1.0\times 10^{18}$ cm$^{-3}$, $\tau_d = 98$ fs, $\sigma_y = 1.15$ $\mu$m, $E_{pk} = 0.5$ GeV, and $\Delta E_{ini} = 2.5 \%$.}
\label{fig_OA_diag}
\end{figure*}

In experimental setups, it is common for the laser propagation axis and the external beam axis to be slightly misaligned. In our study, we have investigated the impact of such misalignment (i.e., the off-axis injection of the external beam) on both the acceleration process and the quality of the accelerated beam.

The schematic of the simulation setup for the off-axis (with respect to the laser) injection of the external electron beam is depicted in Fig. \ref{fig_OA_sch}(a). The degree of misalignment is characterized by the transverse offset distance $\Delta y$ between the laser and the beam axis. In this case, the target in the booster stage is the same as in the previous setup, with an initial plasma density at the plateau of $n_0 = 1.0 \times 10^{18}$ cm$^{-3}$. The parameters for the laser and external beam are also identical to those specified in Table \ref{table1}.

Snapshots of the electron density for both the external beam and the plasma target in the booster stage are shown in Fig. \ref{fig_OA_sch}(b)-(e) for a specific case with $\Delta y = 3.0$ $\mu$m. These snapshots reveal that the external beam is successfully injected into the bubble created by the laser pulse. As long as the offset distance $\Delta y$ is significantly smaller than the bubble radius ($R \sim w_0$), the charge coupling process remains unaffected. Moreover, as the beam travels through the plasma medium, it exhibits oscillatory motion in the transverse plane within the wake cavity. This oscillation is due to the radial component ($E_{wr}$) of the wake electric field. Such oscillatory motion was not observed in the case of on-axis injection (see Fig. \ref{fig_n0td}), as $E_{wr}$ is zero along the laser propagation axis and has a radial symmetry. 

As the entire electron beam undergoes oscillatory motion, the angle between the beam and the laser’s propagation axis ($y = 0$) varies sinusoidally with the propagation distance. This phenomenon is explicitly illustrated in Fig. \ref{fig_OA_diag}(a) for a specific simulation, where the beam has an initial transverse offset of $\Delta y = 3.0$ $\mu$m relative to the laser axis. The divergence angle ($\theta$) of individual particles is calculated using the expression $\theta = |\tan^{-1}{(P_{yi}/P_{xi})}|$. Consequently, the phase of the oscillatory motion (i.e., the angle between the beam and the propagation axis) at the exit will depend on the plasma density, which determines the oscillation wavelength, as well as the length of the plasma channel.

The characteristic properties of the beam, analyzed as a function of propagation distance for different values of $\Delta y$, are shown in Fig. \ref{fig_OA_diag}(b)-(c). Both the relative energy spread and the peak energy remain unaffected by $\Delta y$, as demonstrated in Fig. \ref{fig_OA_diag}(b) and its inset. The trends in relative energy spread ($\Delta E_r$) with propagation distance and the final peak energy gained by the external beam are consistent with those observed in the on-axis injection case. Therefore, it can be concluded that the acceleration process is unaffected by off-axis injection. However, the normalized beam emittance is significantly affected by off-axis injection, as shown in Fig. \ref{fig_OA_diag}(c). After a certain propagation distance, $\epsilon_n$ begins to increase, with a rate being higher for larger values of $\Delta y$. It is important to note that in these cases, the initial transverse beam size is approximately the same as the matched beam size, $\sigma_{ym} \approx 1.2$ $\mu$m. The observed increase in emittance in Fig. \ref{fig_OA_diag}(c) is attributed to the transverse oscillatory motion (i.e., betatron oscillation) of the entire external beam caused by off-axis injection \cite{glinec2008direct}. Larger $\Delta y$ results in greater amplitude of transverse oscillations, leading to a higher rate of emittance growth. Notably, $\epsilon_n$ starts to increase only after a certain propagation distance, which is related to the time required for the phase mixing of the transverse oscillations of the individual electrons in the beam.

\section{Summary and Conclusions}
\label{summary}

Laser wakefield acceleration (LWFA) using the external injection scheme has been explored through two-dimensional Particle-In-Cell (PIC) simulations. This study models a multi-stage LWFA configuration, where electrons accelerated in an initial stage are further accelerated in a subsequent booster stage. Special attention is given to the interaction between externally injected electrons and the wake electric field in the booster stage. Both qualitative and quantitative analyses are provided, highlighting how charge coupling and acceleration are influenced by various system parameters. For instance, the efficiency of beam injection into the focusing and accelerating phase of the wakefield can be independently controlled by adjusting the plasma density in the booster stage and the timing of beam injection relative to the laser driver. A dimensionless parameter $\Lambda$ has been defined as a function of plasma density, laser pulse duration, laser intensity, and offset timing of injection of the external beam with respect to the driver laser, which predicts the charge coupling efficiency in the booster stage. We have shown that the charge coupling efficiency becomes almost a hundred percent for some parametric regimes for which the parameter $\Lambda >0$. In this case, the net beam charge remains constant throughout the evolution. Thus, the final charge coupling efficiency also remains nearly one hundred percent. For $\Lambda<0$, a significant part of the electron beam is injected into the decelerating and defocusing phase of the wakefield, which makes the initial charge coupling efficiency much less than a hundred percent. Additionally, as it evolves, electrons not injected into the accelerating and focusing phase of the wakefield scatter away in the transverse direction, decreasing beam charge as it propagates. Thus, the final charge coupling efficiency decreases further. The study thoroughly examines the evolution of externally injected beam properties, including peak energy, energy spread, and beam emittance, during acceleration in the booster stage. Key parameters affecting the quality of the accelerated beam are identified. In particular, we have demonstrated that plasma density has a significant impact on both the energy gain and energy spread of the beam. Notably, the normalized beam emittance remains independent of plasma density, provided that the entire beam is successfully injected into the accelerating phase of the wakefield. Our study also highlights that the temporal offset of beam injection significantly influences the charge coupling efficiency and the overall quality of the accelerated beam. Thus, the timing synchronization between the successive stages is essential for an efficient multi-stage LWFA setup. Our findings indicate that initial beam characteristics, such as beam density and initial energy spread, do not influence the charge coupling process in the booster stage. Additionally, the initial transverse size of the beam has no effect on the injection and coupling process, as long as it remains significantly smaller than the laser spot size. However, both the initial beam density and transverse size do impact the final energy gain and energy spread due to the space-charge effect. While the normalized beam emittance remains independent of initial beam density and energy spread, it is significantly affected by the initial transverse size of the beam through a phenomenon known as betatron phase-mixing. Additionally, the study investigates the effects of off-axis injection on the acceleration process and the resultant beam properties. Our findings reveal that even for an initially matched beam, a slight off-axis injection (on the order of $3.0$ $\mu$m) can lead to a significant emittance growth. The insights gained from this research are valuable for advancing multi-stage LWFA experiments.


\section*{Acknowledgments}

This work was supported by the Ministry of Education, Youth and Sports of the Czech Republic through the e-INFRA CZ (ID:90254). The authors acknowledge ELI Beamlines HPC facility for computational resources.
 
 

\section*{References}
\bibliographystyle{unsrt}
\bibliography{ref.bib}

\end{document}